\documentclass[preprint,twocolumn
]{aastex631}
\pdfoutput=1

\newcommand{\kms}{$\mathrm{km\,s^{-1}}$}

\begin{document}

\title{Laboratory and astronomical discovery of magnesium dicarbide, MgC$_2$}

\author{P. B. Changala}
\affiliation{Center for Astrophysics $|$ Harvard \& Smithsonian, Cambridge, MA 02138, USA}

\author{H. Gupta}
\affiliation{Center for Astrophysics $|$ Harvard \& Smithsonian, Cambridge, MA 02138, USA}
\affiliation{National Science Foundation, Alexandria, VA 22314, USA}

\author{J. Cernicharo}
\affiliation{Instituto de F\'{i}sica Fundamental, CSIC, Department of Molecular Astrophysics, Serrano 121, 28006 Madrid, Spain}

\author{J. R. Pardo}
\affiliation{Instituto de F\'{i}sica Fundamental, CSIC, Department of Molecular Astrophysics, Serrano 121, 28006 Madrid, Spain}

\author{M. Ag\'{u}ndez}
\affiliation{Instituto de F\'{i}sica Fundamental, CSIC, Department of Molecular Astrophysics, Serrano 121, 28006 Madrid, Spain}

\author{C. Cabezas}
\affiliation{Instituto de F\'{i}sica Fundamental, CSIC, Department of Molecular Astrophysics, Serrano 121, 28006 Madrid, Spain}

\author{B. Tercero}
\affiliation{Observatorio Astron\'{o}mico Nacional, IGN, C/Alfonso XII 3, 28014 Madrid, Spain}

\author{M. Gu\'{e}lin}
\affiliation{Institut de Radioastronomie Millim\'{e}trique, 300 rue de la Piscine, 38406 Saint Martin d'H\`{e}res, France}

\author{M. C. McCarthy}
\affiliation{Center for Astrophysics $|$ Harvard \& Smithsonian, Cambridge, MA 02138, USA}

\begin{abstract}
\noindent
We report the detection of magnesium dicarbide, MgC$_2$, in the laboratory at centimeter wavelengths and assign $^{24}$MgC$_2$, $^{25}$MgC$_2$, and $^{26}$MgC$_2$ to 14 unidentified lines in the radio spectrum of the circumstellar envelope of the evolved carbon star IRC+10216. The structure of MgC$_2$ is found to be T-shaped with a highly ionic bond between the metal atom and the C$_2$ unit, analogous to other dicarbides containing electropositive elements. 
A two-temperature excitation model of the MgC$_2$ emission lines observed in IRC+10216 yields a very low rotational temperature of $6\pm1$~K, a kinetic temperature of $22\pm13$~K, and a column density of $(1.0 \pm 0.3) \times 10^{12}$~cm$^{-2}$.
The abundance of MgC$_2$ relative to the magnesium-carbon chains MgCCH, MgC$_4$H, and MgC$_6$H is $1{:}2{:}22{:}20$ and provides a new constraint on the sequential radiative association-dissociative recombination mechanisms implicated in the production of metal-bearing molecules in circumstellar environments.

\end{abstract}


\section{Introduction}

Silicon dicarbide (SiC$_2$) was long known to exist in carbon-rich stellar atmospheres~\citep[and references therein]{Kleman1956}, but its identification as the carrier of nine unassigned lines in the radio spectrum of the archetype carbon star IRC+10216 was a key advance in  understanding the molecular constituents of evolved carbon stars~\citep{Thaddeus1984}.  Its substantial abundance and the highly ionic nature of the Si--C$_2$ bond suggested that molecules containing other electropositive elements bonded to C$_2$ or longer carbon chains might be present in IRC+10216 and responsible for other unidentified lines there.  In the nearly 40 years since the radio astronomical identification of SiC$_2$, a number of metal-bearing molecules have been discovered in IRC+10216, most containing strong ionic bonds between metal atoms and carbon chains~\citep{Cernicharo2019,Pardo2021}.  It is surprising, therefore, that metal dicarbides  --- the simplest molecules containing one metal atom and two carbon atoms --- are absent from this inventory.  Although several metal dicarbides have now been studied in the laboratory, accurate data are available only for one (AlC$_2$)~\citep[\& references therein]{Halfen2018} that is a plausible candidate for radio astronomical detection.

Magnesium dicarbide (MgC$_2$) is an excellent candidate for detection in IRC+10216 because of the high cosmic abundance of magnesium, which is found in nearly half of all known metal-bearing species in this source. 
An early theoretical study of MgC$_2$~\citep{Green1984} predicted a triangular, symmetric, and highly polar ground-state structure similar to SiC$_2$.  Motivated by the discoveries of MgNC and MgCN several years later~\citep{Kawaguchi1993,Ziurys1995}, more accurate calculations of MgC$_2$ appeared in the literature~\citep{Itono2000}, but no gas-phase measurements followed.  The purpose of this Letter is to report the laboratory detection of MgC$_2$ at centimeter wavelengths, and its assignment to a series of previously unidentified lines in the radio spectrum of IRC+10216.  To our knowledge, this is the first spectroscopic characterization of MgC$_2$ in any wavelength region, while its astronomical discovery  fills a longstanding gap in the molecular inventory of evolved carbon stars.

\section{Laboratory measurements}

Rotational lines of MgC$_2$ were detected with a Fourier transform microwave (FTMW) spectrometer coupled to a laser ablation supersonic expansion source previously used in our laboratory to characterize small metal-containing molecules~\citep{Brunken2008,Zingsheim2017,Lee2019}. A rotating magnesium rod was ablated by 50-mJ pulses of 532-nm radiation from an Nd:YAG laser operating at 5~Hz. The ablated Mg atoms were entrained in a dilute mixture (0.1\%) of acetylene in neon, which then passed through two copper discharge electrodes. An 800-V discharge struck between the electrodes increased the product yield by up to a factor of 5~\citep{Sun2010}. After expanding into a large vacuum chamber, the molecules were polarized in a co-axially aligned cavity spectrometer operating from 5 to 26~GHz~\citep{Grabow2005,Crabtree2016}. Double-resonance transitions at higher frequencies were driven by microwave horns placed perpendicular to the cavity axis.

\begin{figure}[h]
\centering
\includegraphics[width=3.2in]{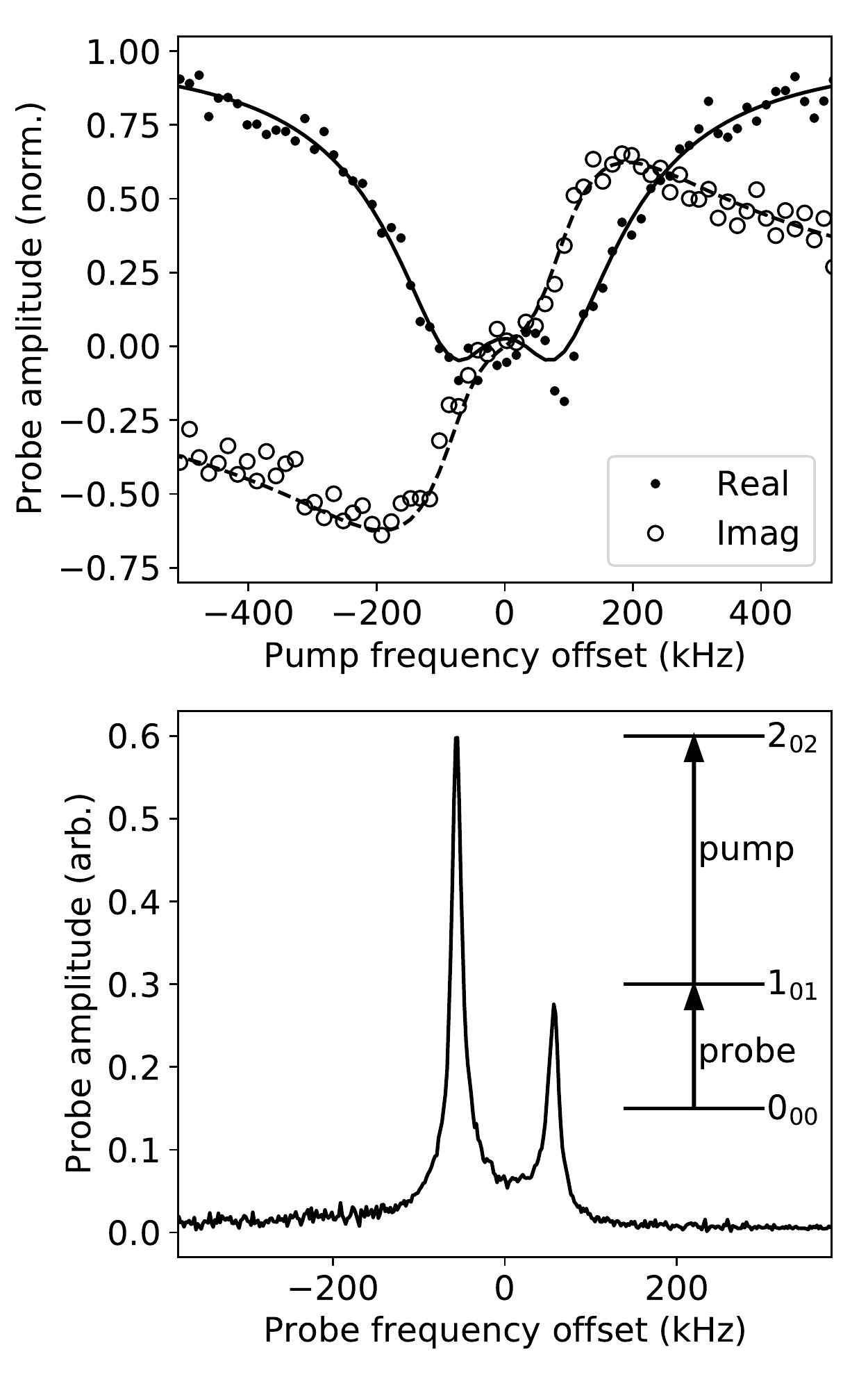}
\caption{\label{fig:cavity_spec}Laboratory measurements of the two lowest rotational transitions of $^{24}$MgC$_2$. The cavity-FTMW signal (bottom) centered at 20896.090~MHz is split into two Doppler components due to the co-axial cavity geometry. Its complex amplitude response (top) is plotted as a function of the double resonance pump frequency relative to 41711.852~MHz. The solid and dashed curves are the fitted saturated Lorentzian line profiles for the real and imaginary parts, respectively. Each panel represents 6.5~h of integration or about 120,000 laser shots.}
\end{figure}

The identification of MgC$_2$ was challenging because only one transition, the $1_{01}-0_{00}$ line predicted to lie near 20.9~GHz~\citep{Itono2000}, falls within our spectrometer bandwidth. 
After optimizing the production of MgCCH, we conducted a search from 20.4 to 21.0~GHz. This survey was extremely sparse, with only a single ablation-dependent feature at 20896~MHz, shown in Fig.~\ref{fig:cavity_spec}, close to the prediction of 20892~MHz from quantum chemical calculations performed for this work (see Appendix~\ref{sec:calc}). 
We conducted a number of assays to test the carrier of this line: the signal was unaffected by applied magnetic fields, consistent with the predicted closed-shell ground electronic state; it required the presence of acetylene, but was unaffected by replacing HCCH with DCCD, indicating it contains carbon but not hydrogen; and the relative intensity from a statistical $^{12}$C/$^{13}$C HCCH precursor suggested it likely contained only two C atoms.
Spectroscopic confirmation was provided by the observation of a double resonance transition ($2_{02}-1_{01}$, top panel of Fig~\ref{fig:cavity_spec}) at its expected frequency near 41.7~GHz.
Assuming a dipole moment of 7.9~D~\citep{Itono2000}, we estimate the number of MgC$_2$ molecules in each gas pulse to be $4 \times 10^8$, significantly less abundant than MgCCH ($5 \times 10^{10}$) or the precursor Mg atoms ($5 \times 10^{14}$, estimated from the ablation target mass loss) and HCCH ($2\times 10^{15}$, from calibrated flow meters).

With the assignment of MgC$_2$ secure, we proceeded to search for its isotopologues. The fundamental transitions of $^{26}$Mg$^{12}$C$_2$ and $^{25}$Mg$^{12}$C$_2$ were detected in natural abundance,
while those of $^{24}$Mg$^{13}$C$^{12}$C and $^{24}$Mg$^{13}$C$_2$ were detected using a $^{13}$C-enriched sample of acetylene. Their transition frequencies are shown in Table~\ref{tab:lab_data} and are uniformly within 0.5~MHz of the theoretical isotopic shifts. These laboratory measurements are insufficient to determine the $A$, $B$, and $C$ rotational constants for each isotopologue independently, but together provide enough information to derive a semi-experimental equilibrium structure, shown in Fig.~\ref{fig:rotation_levels} (see Appendix~\ref{sec:calc} for details). The $1_{01}-0_{00}$ transition of $^{25}$Mg$^{12}$C$_2$ is split into three hyperfine components by nuclear electric quadrupole interactions. The observed coupling constants are $\chi_{aa} = -11.75(2)$~MHz and $\chi_{bb} = 4.8(10)$~MHz, which agree well with the theoretical predictions of $-11.4$~MHz and $4.8$~MHz, respectively.

\section{Astronomical identification}

With accurate laboratory data in hand, we examined available spectroscopic surveys and observations of IRC+10216 for evidence of MgC$_2$.  An unidentified rotational line, U41712, coincident with the $2_{02}-1_{01}$ transition of $^{24}$MgC$_2$ was found in the recent 6--10~millimeter survey done by \citet{Pardo2022} using the Yebes~40~m telescope.  A second unidentified line, U82788, was found in a study reporting the identification of the HC$_4$N radical~\citep{Cernicharo2004} using the IRAM~30~m telescope.  Subsequently, 8 additional unidentified lines were quickly found in unpublished surveys at 2 and 3 millimeters done with the IRAM~30~m telescope (see~\citet{Cernicharo2019} for details) and assigned to $^{24}$MgC$_2$.  One line of $^{25}$MgC$_2$ and three lines of $^{26}$MgC$_2$ were also found at precisely the expected isotopic shifts.

\begin{figure*}[h]
\centering 
\includegraphics[width=1.0\textwidth]{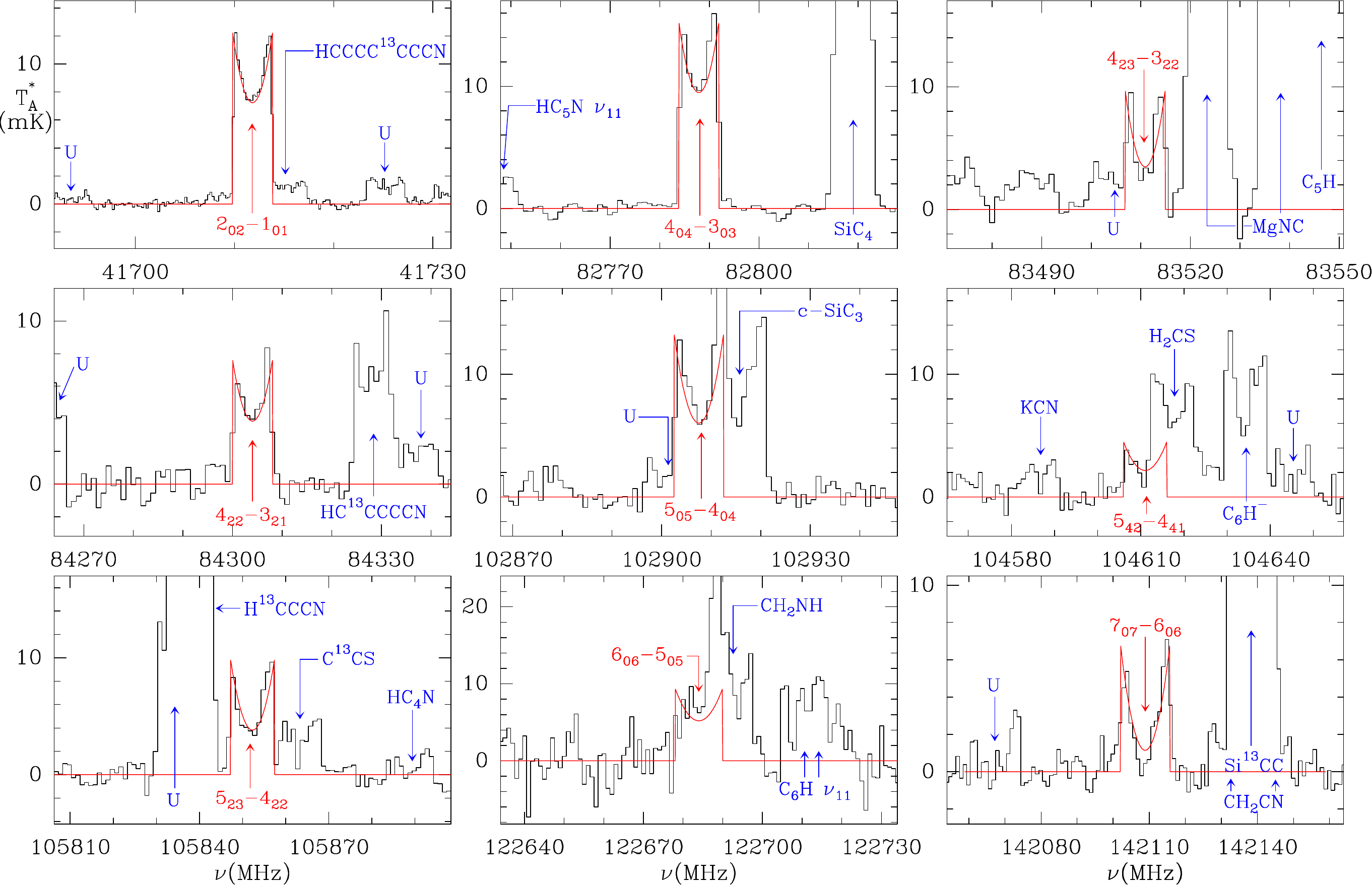}
\caption{\label{fig:astro_main} Spectra of $^{24}$MgC$_2$ toward IRC+10216. The fitted profiles are plotted in red. `U' indicates unidentified lines. }
\end{figure*}

\begin{figure*}[h]
\centering 
\includegraphics[width=1.0\textwidth]{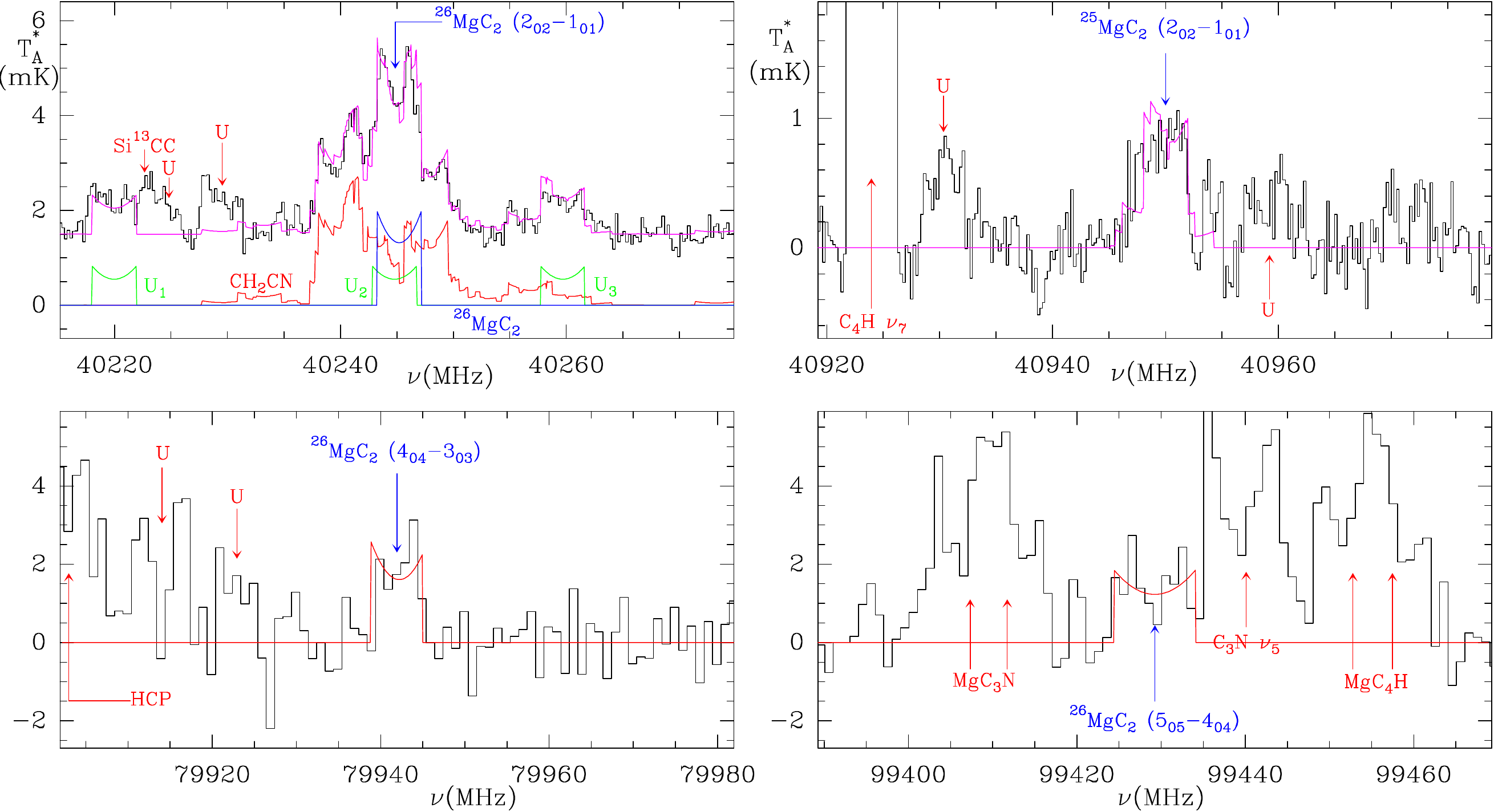}
\caption{\label{fig:astro_isotope} Spectra of $^{25}$MgC$_2$ and $^{26}$MgC$_2$ toward IRC+10216. The simulated profile of the $2_{02} - 1_{01}$ transition of $^{26}$MgC$_2$ (shown in blue in the top left panel) accounts for contamination from overlapping CH$_2$CN transitions (shown in red); the total emission is shown in magenta. The profile for $^{25}$MgC$_2$ (top right) includes quadrupole hyperfine splittings calculated from the laboratory constants.
`U' indicates unidentified lines.}
\end{figure*}

There is overwhelming evidence that MgC$_2$ is the carrier of the lines listed in Table~\ref{tbl-2} and displayed in Figures~\ref{fig:astro_main} and \ref{fig:astro_isotope}. As Table~\ref{tbl-2} shows, a combined fit of the observed laboratory and astronomical lines of $^{24}$MgC$_2$ to Watson's $A$-reduced Hamiltonian yields small residuals corresponding to a normalized rms of 1.37 (Table~\ref{tbl-3}), demonstrating that they originate from the same molecule. Table~\ref{tbl-3} lists the spectroscopic constants derived using only six free parameters (three rotational constants and three centrifugal distortion constants) and two parameters fixed to theoretical values.
Similarly small residuals are obtained for $^{25}$MgC$_2$ and $^{26}$MgC$_2$, albeit with fewer determinable parameters given the small data sets.
The ratio of the integrated intensities of the $2_{02}-1_{01}$ lines of the three isotopologues is $(73\pm5)\rm{:}(11\pm2)\rm{:}(15\pm3)$, consistent with the solar abundance ratio of $^{24}$Mg:$^{25}$Mg:$^{26}$Mg of $(79.2\pm0.6)\rm{:}(9.5\pm0.5)\rm{:}(11.3\pm0.5)$~\citep{Bochsler1996} and the abundance ratio derived for MgNC in IRC+10216 itself, $(78\pm2)\rm{:}(11\pm1)\rm{:}(11\pm1)$~\citep{Guelin1995}.

Figures~\ref{fig:astro_main} and \ref{fig:astro_isotope} present the astronomical spectra and show that the assigned lines meet all requirements of the identification.  
Each line exhibits the same double-peaked or ``U-shaped'' profile and has the same width as the strong $4_{04}-3_{03}$ line ($29.0\pm 0.5$~\kms) to within observational uncertainties. Six fully resolved or nearly fully resolved lines and four partially blended lines of $^{24}$MgC$_2$ are detected. Three lines of $^{26}$MgC$_2$ are detected, including one which is partially blended. Finally, one line of $^{25}$MgC$_2$ is detected with a slightly broadened line width consistent with the underlying quadrupole hyperfine structure predicted from the laboratory measurements.  The high quality profiles derived from the resolved lines and the precise rest frequencies of the blended lines allow us to account for the contributions of the interlopers and extract reliable parameters.

We have analyzed the intensities of the rotational lines to infer the excitation of the molecules on two assumptions: (1) to correct for the changing antenna beamwidth in Table~\ref{tbl-2}, it is assumed by analogy with other Mg-bearing molecules in IRC+10216, that the MgC$_2$ emission is confined to a shell of $30\arcsec$ in diameter~\citep{Cernicharo2019,Pardo2021}, an assumption supported by the U-shaped profiles of the lines and (2) all of the lines in Table~\ref{tbl-2} are optically thin because of their faintness.

\section{Discussion}

\begin{figure}[t]
\centering 
\vspace{3mm}
\includegraphics[width=0.45\textwidth]{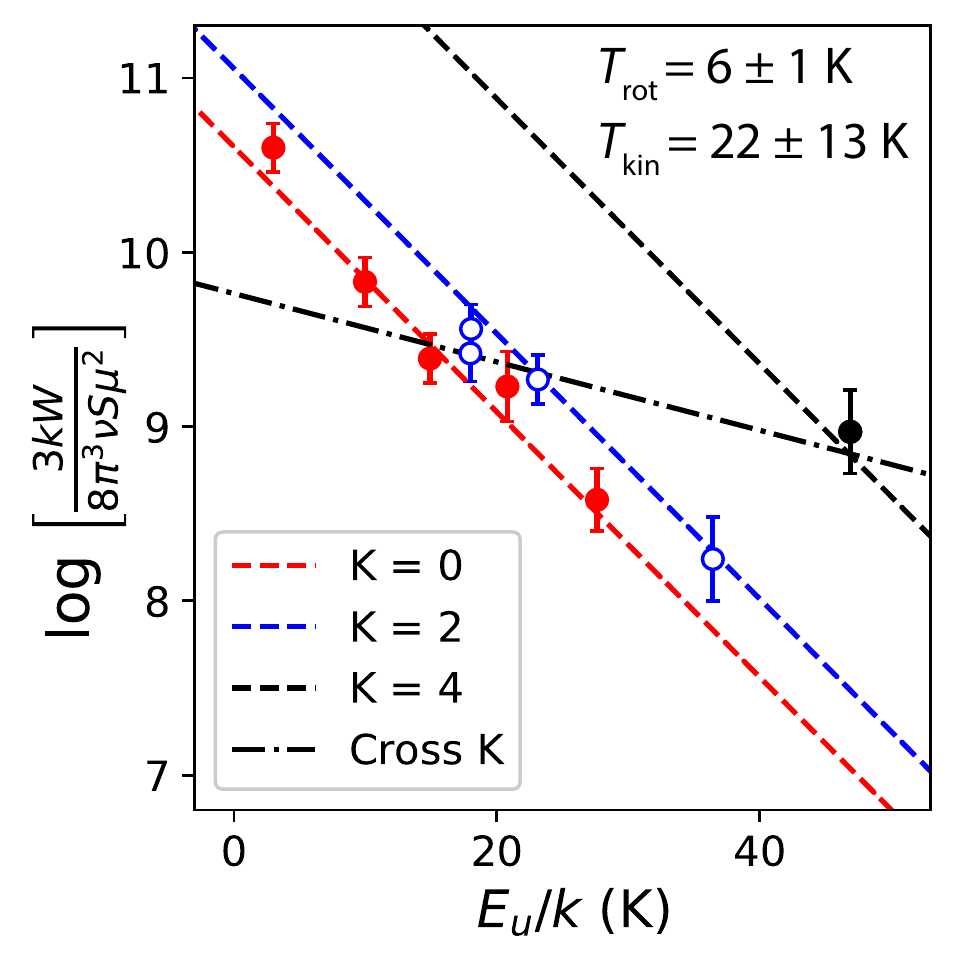}
\caption{\label{fig:RotDiag} Rotational temperature diagram of $^{24}$MgC$_2$ in IRC+10216, fitted  with the two-temperature model described in Appendix~\ref{sec:RotEx}.  The slope of the parallel lines through each $K$ ladder yields the rotational temperature ($T_\mathrm{rot}$), and the slope of the line through points across the $K$ ladders yields the kinetic temperature ($T_\mathrm{kin}$).  The flux of the $K=4$ transition is assumed to be one-half the total flux of the unresolved $5_{42} - 4_{41}$/$5_{41} - 4_{40}$ doublet. The error bar on each data point is twice the error shown in Table~\ref{tbl-2}.  Error bars on the parameters are twice the standard error of the fit.}
\end{figure}

As Figure~\ref{fig:rotation_levels} shows, rotational emission lines spanning a wide range in energy are detected in IRC+10216, allowing us to construct the rotational temperature diagram shown in Fig.~\ref{fig:RotDiag}.  
The integrated intensities of the lines are reproduced by a simple two-temperature excitation model (see Appendix~\ref{sec:RotEx})
in which a temperature $T_\mathrm{rot}=6\pm1$~K describes the relative populations within each $K$ ladder
and $T_\mathrm{kin}=22\pm13$~K describes the relative populations across $K$ ladders and provides a measure of the gas kinetic temperature. 
A comparison of the derived excitation temperatures of MgC$_2$, SiC$_2$, and SiC$_3$ shows that $T_\mathrm{rot}$ is similar for the three species ($6\pm1$~K vs. 10~K and 13~K~\citep{Thaddeus1984,Apponi1999}), but $T_\mathrm{kin}$ is not.  The latter is similar for MgC$_2$ and SiC$_3$ ($22\pm13$~K and 46~K~\citep{Apponi1999}), and it is significantly smaller than for SiC$_2$ (140~K~\citep{Thaddeus1984}), probably because MgC$_2$ and SiC$_3$ reside in the cool outer envelope of IRC+10216, whereas SiC$_2$ is more widely distributed, residing in the warmer inner regions as well as the outer envelope~\citep{Patel2011}. The $T_\mathrm{kin}$ of 22~K inferred for MgC$_2$ is remarkably close to the value of 20~K inferred from observations and modeling of dust, CO, and CCH in the $30\arcsec\rm{-}40\arcsec$ diameter shell of IRC+10216~\citep{Guelin2018}.  As discussed in Appendix~\ref{sec:RotEx}, the column density averaged over a $30\arcsec$ antenna beam is  $N(\mathrm{MgC_{2}})=1\times10^{12}$~cm$^{-2}$, with an estimated uncertainty of $30\%$, corresponding to a fractional abundance of ${\sim}5\times10^{-10}$ relative to H$_2$.\footnote{We adopt $N(\mathrm{H_{2}})=2.1\times10^{21}$~cm$^{-2}$ following \citet{Gong2015}.}

On the assumption that MgC$_2$ is co-spatial with the magnesium-carbon chain radicals MgCCH, MgC$_4$H and MgC$_6$H (which is supported by the similar line profiles), a comparison of their abundances might shed light on the production of Mg-bearing species in the outer envelope of IRC+10216. The formation of MgCCH, MgC$_4$H and MgC$_6$H has been explained by a two-step process in which radiative association (RA) of Mg$^+$ with large polyynes HC$_{2n}$H ($n\geq4$) yields the intermediate ions MgHC$_{2n}$H$^{+}$, which then undergo dissociative recombination (DR) to yield the neutral radicals~\citep{Cernicharo2019,Pardo2021}. 
This mechanism, originally proposed by \citet{Petrie1996} and later expanded by \citet{Dunbar2002}, is thought to control the formation of the metal cyanides and acetylides detected in IRC+10216 \citep{Millar2008,Cabezas2013,Cernicharo2019,Pardo2021}. The chemical models developed in these studies indicate that the main precursors of magnesium carbon chains are large MgHC$_{2n}$H$^+$ complexes with more than six carbon atoms.
If, as is plausible,  MgC$_2$ is also a fragmentation product in the DR of MgHC$_{2n}$H$^{+}$, then the observed column density ratio MgC$_2$:MgCCH:MgC$_4$H:MgC$_6$H~$=1{:}2{:}22{:}20$ provides a useful constraint on the relative yields of the DR of the parent ions. Observations of other plausible DR products, such as MgC and larger MgC$_n$ clusters are needed to better elucidate the poorly constrained gas-phase chemistry. A parallel formation mechanism that may also be important is the RA-DR of Mg$^+$ and large cyanopolyynes, HC$_{2n+1}$N~\citep{Dunbar2002}. Experimental measurements of the branching ratios of the various fragments formed in the DR of large MgHC$_{2n}$H$^+$ and MgHC$_{2n+1}$NH$^+$ ions would enable a direct comparison against the observed abundance ratios, permitting a holistic assessment of the production of metal-bearing molecules in the outer layers of IRC+10216. 

Understanding the structure and bonding in metal dicarbides is of fundamental chemical interest and may identify additional metal carbides that are candidates for astronomical detection.
Like other T-shaped triatomic dicarbides, including SiC$_2$~\citep{Cernicharo1991}, GeC$_2$~\citep{Zingsheim2017}, BeC$_2$~\citep{Green2020}, AlC$_2$~\citep{Halfen2018}, ScC$_2$~\citep{Min2014}, and YC$_2$~\citep{Halfen2013}, MgC$_2$ exhibits an ionic metal-carbon bond best described as Mg$^+$--C$_2^-$.
Its equilibrium CC bond length, $r_\mathrm{CC} = 1.2706(7)$~\AA, is close to that of bare C$_2^-$ itself, $r_\mathrm{CC} = 1.26831(13)$~\AA~\citep{Rehfuss1988}, which are both significantly longer than neutral C$_2$, $r_\mathrm{CC} = 1.24244(1)$~\AA~\citep{Douay1988}. Moreover, the ionicity, defined as the ratio of the molecular dipole moment to the dipole moment resulting from unit point charges separated by the metal-C$_2$ bond length, $i = \mu/er_\mathrm{Mg-C_2} = 0.86$, suggests nearly complete transfer of a $3s$ electron from Mg to the C$_2$ fragment.

The bonding in MgC$_2$ closely resembles that in MgO~\citep{Boldyrev1997}, with the in-plane $\pi_u$ orbital of C$_2$ taking the place of the $2p_z$ orbital of O. MgO has a similar ionicity, $i = 0.72$, and Mg nuclear quadrupole coupling constant: $-10(4)$~MHz~\citep{Torring1986} versus $\chi_{aa} = -11.72(2)$~MHz for MgC$_2$. These latter parameters indicate that the Mg $3s$ orbital in both molecules is substantially hybridized. Given that the atomic $3p_z$ orbital of Mg$^{+}$ has a quadrupole coupling constant of $-46$~MHz~\citep{Sur2005}, the $p$ character is about 25\%. Although simple orbital descriptions of MgC$_2$ and MgO are potentially obscured by their multiconfigurational, mixed ionic-covalent character~\citep{Thummel1989,Itono2000}, this issue appears to be less severe for MgC$_2$, which is adequately described with single-reference coupled cluster methods as borne out by the calculations in this work and others~\citep{Woon1996}.
The bonding of the heavier Group IIA dicarbides CaC$_2$ and SrC$_2$ can be expected to follow the same trends observed in their respective oxides. Because of their high ionicities, all of these species possess unusually large dipole moments~\citep{Fuentealba2000}.

Observations of MgC$_2$ and similar metal dicarbides are a promising means to elucidate the role of metals in the chemistry of IRC+10216, as well as the state of refractory elements in carbon-rich circumstellar environments more generally.  Metals are inferred to be significantly depleted onto dust grains in IRC+10216~\citep{Mauron2010}, yet even the residual gas-phase abundance of metal atoms and ions is sufficient to drive a rich chemistry. For example, molecules containing the cyano radical (CN) bonded to the most abundant metallic elements (Na, K, Mg, Ca, Al, and Fe) have been found there. Given that the abundances of MgC$_2$ and MgNC~\citep{Guelin1995} are within an order of magnitude of each other, it is very likely that other metal dicarbides are also synthesized in the external layers of the circumstellar envelope of IRC+10216, plausibly following the same chemical pathway that is postulated to yield MgC$_2$. By analogy with SiC$_2$, which has yielded invaluable information on gas-phase processes within the inner and intermediate winds, as well as the outer circumstellar envelopes of evolved carbon-rich stars~\citep[and references therein]{Massalkhi2018}, metal dicarbides may also serve as important probes provided their excitation, abundance, and distribution are understood.  The distribution of metal dicarbides as a function of radial distance from the central star in IRC+10216, for example, accompanied by a careful analysis of their abundance and excitation might shed light on processes such as condensation and dust formation, which remain poorly understood.

The identification of MgC$_2$ in the laboratory and in space marks a key step in the study of metal carbides and suggests that other highly polar metal-carbon molecules might be detectable.  After MgC$_2$, CaC$_2$ is the next most promising candidate for detection because it possesses an even larger dipole moment (10.7~D;~\citet{Fuentealba2000}) than MgC$_2$  (7.9~D;~\citet{Itono2000}) and because Ca is cosmically the next most abundant Group~IIA element after Mg; calcium isocyanide (CaNC) has already been detected in IRC+10216~\citep{Cernicharo2019}, suggesting that other Ca-bearing molecules may be present there. By analogy with SiC~\citep{Massalkhi2018}, the monocarbides MgC and CaC may be important for comparative studies of the distribution, abundance, and photochemical processing of dicarbides in circumstellar envelopes; although CaC has been detected in the laboratory, MgC has not, and its astronomical detection might be more feasible given the higher abundance of Mg.  Other promising systems for study include larger metal-carbon clusters such as MgC$_{n}$ about which little is known experimentally, but which, like the silicon-carbon clusters SiC$_3$ and SiC$_4$ \citep{Apponi1999,Gordon2000}, are amenable to laboratory characterization and astronomical detection.

\pagebreak

\appendix

\section{Structure calculations and semi-experimental equilibrium geometry}\label{sec:calc}

Coupled cluster calculations were performed using the CFOUR package~\citep{Matthews2020} at the singles, doubles, and perturbative triples (CCSD(T)) level of theory~\citep{Raghavachari1989} with the cc-pCV$X$Z ($X$ = D,T,Q,5) basis sets~\citep{Woon1995,Prascher2011} correlating all but the Mg $1s$ electrons. The complete basis set equilibrium geometry was estimated by an exponential extrapolation of the entire $X$ = D--5 sequence. To derive a semi-experimental equilibrium geometry, vibrational corrections to the rotational frequencies were determined by variational rovibrational calculations on a  potential energy surface (PES) fitted to 448 CCSD(T)/cc-pCVQZ single-point energies using the NITROGEN package~\citep{NITROGEN2021}. Additional contributions to the rotational constants from the rotational $g$-tensor~\citep{Gauss2010} were computed with the cc-pCVTZ basis set, and nuclear electric quadrupole coupling constants with the cc-pCV5Z basis set. 

The small number of laboratory transitions observed for each isotopologue is insufficient to determine a complete set of experimental rotational constants. Instead, the structure determination was performed by least-squares fitting directly to the transition frequencies. That is, a proposed equilibrium geometry was used to compute equilibrium rotational constants, which were then corrected by the rotational $g$-tensor. These in turn were used to generate rigid-rotor transition frequencies, to which the vibrational corrections from the rovibrational variational calculations were added. These latter corrections are defined as the difference between the variational transition energies and the rigid-rotor transition energies based on the equilibrium geometry of the same PES. They therefore include both vibrational zero-point and centrifugal distortion effects. For $^{25}$MgC$_2$, the degeneracy-weighted mean of the quadrupole-split energies was used as the hyperfine-free value. The optimized structural parameters are shown in Fig.~\ref{fig:rotation_levels} and compared to the extrapolated CCSD(T) geometry in Table~\ref{tab:geometry}. These parameters reproduce the 6 laboratory frequencies with a root-mean-square residual of 0.1~MHz. This is significantly larger than the experimental uncertainty (2~kHz) indicating that the theoretical corrections are the dominant source of error.

%\clearpage
%\pagebreak

\section{Two-temperature model for rotational excitation}\label{sec:RotEx}

In the analysis of rotational emission of prolate asymmetric tops such as MgC$_2$ the partitioning of energy levels within and across radiatively decoupled rotational manifolds must be taken into account.  Because radiative transitions are confined within the $K$ ladders but collisional transitions occur within and across the ladders, the rotational excitation temperature within the ladders ($T_\mathrm{rot}$) is not necessarily equal to the excitation temperature across the ladders, which is assumed equal to the kinetic temperature ($T_\mathrm{kin}$) of the gas.  For highly polar species, excitation within the ladders is highly subthermal (i.e., $T_\mathrm{rot} \ll T_\mathrm{kin}$) because collisional excitation is much slower than spontaneous radiative decay.  Let $T_\mathrm{kin}$ describe the population distribution across $K$ ladders and $T_\mathrm{rot}$ describe the population distribution within each $K$ ladder.  Then, the intensity of an optically thin rotational line is given by
\begin{equation}\label{eq:B1}
   \frac{3kW}{8\pi^{3}\nu S \mu^{2}}= \frac{Ne^{-E_{J}/kT_\mathrm{rot}}e^{{-E_{K}/kT_\mathrm{kin}}}}{Z}   
\end{equation}
where $W=\int{T_\mathrm{MB}\,dv}$ is the main beam temperature integrated over the radial velocity, $\nu$ is the line frequency, $S$ is the rotational line strength, $\mu$ is the dipole moment, and $N$ is the total column density. The Boltzmann factors and partition function, $Z$, in Eq.~\ref{eq:B1} are calculated by splitting the total rotational energy into two parts, $E_u = E_{K} + E_{J}$, where $E_{K}=(A-{(B+C)}/{2})K^{2}$ is the $K$-dependent part, and $E_J$ is the remainder. Taking the logarithm of Eq.~\ref{eq:B1} yields
\begin{equation}\label{eq:B2}
  \log \frac{3kW}{8\pi^{3}\nu S \mu^{2}}= \log \frac{N}{Z} - \frac{(E_{u}-E_{K}) \log e}{kT_\mathrm{rot}} - \frac{E_{K} \log e}{kT_\mathrm{kin}}.
\end{equation}
A least squares fit of  equation~\ref{eq:B2} to the integrated intensities ($W$) from Table~\ref{tbl-2} yields $T_\mathrm{rot}=5.71\pm1.12$~K and $T_\mathrm{kin}=22.14\pm13.08$~K, which are shown rounded off in the rotational temperature diagram of Fig.~\ref{fig:RotDiag}.  Multiplying the intercept $\log(N/Z)=10.60$ by the partition function $Z(T_\mathrm{rot},T_\mathrm{kin})=24.59$ evaluated numerically as a sum over states, yields $N(\mathrm{MgC_2})=1\times10^{12}$~cm$^{-2}$. We estimate the total uncertainty in $N$ from the dipole moment ($7.9\pm1.0$~D) and excitation conditions to be 30\%.

\section{Acknowledgments}

We thank Dr. Marie-Aline Martin-Drumel (CNRS) for her valuable assistance with the initial laboratory efforts and comments on the manuscript. P.B.C. and M.C.M. are supported by the National Science Foundation (award nos.  AST-1908576 and PHY-2110489). H.G. acknowledges support from the National Science Foundation for participation in this work as part of his independent research and development plan.  Any opinions, findings, and conclusions expressed in this material are those of the authors and do not necessarily reflect the views of the National Science Foundation.  We acknowledge funding support from the Spanish Ministerio de Ciencia e Innovaci\'{o}n through grants PID2019-107115GB-C21 and PID2019-106110GB-100.

\clearpage
\pagebreak

%\bibliography{references}{}

\begin{thebibliography}{}
\expandafter\ifx\csname natexlab\endcsname\relax\def\natexlab#1{#1}\fi
\providecommand{\url}[1]{\href{#1}{#1}}
\providecommand{\dodoi}[1]{doi:~\href{http://doi.org/#1}{\nolinkurl{#1}}}
\providecommand{\doeprint}[1]{\href{http://ascl.net/#1}{\nolinkurl{http://ascl.net/#1}}}
\providecommand{\doarXiv}[1]{\href{https://arxiv.org/abs/#1}{\nolinkurl{https://arxiv.org/abs/#1}}}

\bibitem[{Apponi {et~al.}(1999)Apponi, McCarthy, Gottlieb, \&
  Thaddeus}]{Apponi1999}
Apponi, A.~J., McCarthy, M.~C., Gottlieb, C., \& Thaddeus, P. 1999, ApJ, 516,
  L103

\bibitem[{Bochsler {et~al.}(1996)Bochsler, Gonin, Sheldon, Th., Gloeckler,
  Hamilton, Collier, \& D.}]{Bochsler1996}
Bochsler, P., Gonin, M., Sheldon, R.~B., {et~al.} 1996, AIP Conference
  Proceedings, 382, 199

\bibitem[{Boldyrev \& Simons(1997)}]{Boldyrev1997}
Boldyrev, A.~I., \& Simons, J. 1997, J. Phys. Chem. A, 101, 2215.
\newblock \url{https://pubs.acs.org/doi/abs/10.1021/jp962907i}

\bibitem[{Br{\"{u}}nken {et~al.}(2008)Br{\"{u}}nken, M{\"{u}}ller, Menten,
  McCarthy, \& Thaddeus}]{Brunken2008}
Br{\"{u}}nken, S., M{\"{u}}ller, H. S.~P., Menten, K.~M., McCarthy, M.~C., \&
  Thaddeus, P. 2008, ApJ, 676, 1367.
\newblock \url{http://www.astro.uni-koeln.de/vorhersagen.}

\bibitem[{Cabezas {et~al.}(2013)Cabezas, Cernicharo, Alonso, Agundez, Mata,
  Guelin, \& Pe{\~{n}}a}]{Cabezas2013}
Cabezas, C., Cernicharo, J., Alonso, J.~L., {et~al.} 2013, 775, 133

\bibitem[{Cernicharo {et~al.}(2019)Cernicharo, Cabezas, Pardo, Ag\'{u}ndez, \&
  Berm\'{u}dez}]{Cernicharo2019}
Cernicharo, J., Cabezas, C., Pardo, J., Ag\'{u}ndez, M., \& Berm\'{u}dez, C.
  e.~a. 2019, Astron. Astrophys., 630, L2

\bibitem[{Cernicharo {et~al.}(2004)Cernicharo, Gu\'{e}lin, \&
  Pardo}]{Cernicharo2004}
Cernicharo, J., Gu\'{e}lin, M., \& Pardo, J.~R. 2004, Astron. Astrophys., 615,
  L145

\bibitem[{Cernicharo {et~al.}(1991)Cernicharo, Gu\'{e}lin, Kahane, Bogey,
  Demuynck, Destombes, Guelin, Kahane, Bogey, \& Demuynck}]{Cernicharo1991}
Cernicharo, J., Gu\'{e}lin, M., Kahane, C., {et~al.} 1991, Astron. Astrophys.,
  246, 213

\bibitem[{Changala(2021)}]{NITROGEN2021}
Changala, P.~B. 2021, {NITROGEN, version 2.1,
  https://github.com/bchangala/nitrogen}

\bibitem[{Crabtree {et~al.}(2016)Crabtree, Martin-Drumel, Brown, Gaster, Hall,
  \& McCarthy}]{Crabtree2016}
Crabtree, K.~N., Martin-Drumel, M.-A., Brown, G.~G., {et~al.} 2016, J. Chem.
  Phys., 144, 124201.
\newblock
  \url{http://scitation.aip.org/content/aip/journal/jcp/144/12/10.1063/1.4944072}

\bibitem[{Douay {et~al.}(1988)Douay, Nietmann, \& Bernath}]{Douay1988}
Douay, M., Nietmann, R., \& Bernath, P.~F. 1988, J. Mol. Spectrosc., 131, 250

\bibitem[{Dunbar \& Petrie(2002)}]{Dunbar2002}
Dunbar, R.~C., \& Petrie, S. 2002, Astrophys. J., 564, 792

\bibitem[{Fuentealba \& Savin(2000)}]{Fuentealba2000}
Fuentealba, P., \& Savin, A. 2000, J. Phys. Chem. A, 104, 10882

\bibitem[{Gauss \& Puzzarini(2010)}]{Gauss2010}
Gauss, J., \& Puzzarini, C. 2010, Mol. Phys., 108, 269

\bibitem[{Gong {et~al.}(2015)Gong, Henkel, \& Spezzano}]{Gong2015}
Gong, Y., Henkel, C., \& Spezzano, S. e.~a. 2015, Astron. Astrophys., 574, A56

\bibitem[{Gordon {et~al.}(2000)Gordon, Nathan, Apponi, McCarthy, Thaddeus, \&
  Botschwina}]{Gordon2000}
Gordon, V.~D., Nathan, E.~S., Apponi, A.~J., {et~al.} 2000, J. Chem. Phys.,
  113, 5311

\bibitem[{Grabow {et~al.}(2005)Grabow, Palmer, McCarthy, \&
  Thaddeus}]{Grabow2005}
Grabow, J.~U., Palmer, E.~S., McCarthy, M.~C., \& Thaddeus, P. 2005, Rev. Sci.
  Instrum., 76, 093106

\bibitem[{Green {et~al.}(2020)Green, Jaffe, \& Heaven}]{Green2020}
Green, M.~L., Jaffe, N.~B., \& Heaven, M.~C. 2020, J. Phys. Chem. Lett., 11, 88

\bibitem[{Green(1984)}]{Green1984}
Green, S. 1984, Chem. Phys. Lett., 112, 29

\bibitem[{Gu{\'{e}}lin {et~al.}(1995)Gu{\'{e}}lin, Forestini, Valiron, Ziurys,
  Anderson, Cernicharo, \& Kahane}]{Guelin1995}
Gu{\'{e}}lin, M., Forestini, M., Valiron, P., {et~al.} 1995, Astron.
  Astrophys., 297, 183

\bibitem[{Gu\'elin {et~al.}(2018)Gu\'elin, Patel, Bremer, Cernicharo,
  Castro-Carrizo, Pety, Fonfr\'{i}a, Ag\'undez, Santander-Garc\'{i}a,
  {Quintana-Lacaci, G.}, {Velilla Prieto, L.}, {Blundell, R.}, \& {Thaddeus,
  P.}}]{Guelin2018}
Gu\'elin, M., Patel, N.~A., Bremer, M., {et~al.} 2018, A\&A, 610, A4

\bibitem[{Halfen {et~al.}(2013)Halfen, Min, \& Ziurys}]{Halfen2013}
Halfen, D.~T., Min, J., \& Ziurys, L.~M. 2013, Chem. Phys. Lett., 555, 31

\bibitem[{Halfen \& Ziurys(2018)}]{Halfen2018}
Halfen, D.~T., \& Ziurys, L.~M. 2018, Phys. Chem. Chem. Phys., 20, 11047

\bibitem[{Itono {et~al.}(2000)Itono, Takano, Hirano, \& Nagashima}]{Itono2000}
Itono, S., Takano, K., Hirano, T., \& Nagashima, U. 2000, ApJL, 538, L163.
\newblock \url{https://ui.adsabs.harvard.edu/abs/2000ApJ...538L.163I/abstract}

\bibitem[{Kawaguchi {et~al.}(1993)Kawaguchi, Kagi, Hirano, Takano, \&
  S.}]{Kawaguchi1993}
Kawaguchi, K., Kagi, E., Hirano, T., Takano, S., \& S., S. 1993, ApJ, 406, L39

\bibitem[{Kleman(1956)}]{Kleman1956}
Kleman, B. 1956, ApJ, 123, 162

\bibitem[{Lee {et~al.}(2019)Lee, Thorwirth, Martin-Drumel, \&
  McCarthy}]{Lee2019}
Lee, K. L.~K., Thorwirth, S., Martin-Drumel, M.~A., \& McCarthy, M.~C. 2019,
  Phys. Chem. Chem. Phys., 21, 18911.
\newblock \url{https://pubs.rsc.org/en/content/articlehtml/2019/cp/c9cp03607e
  https://pubs.rsc.org/en/content/articlelanding/2019/cp/c9cp03607e}

\bibitem[{Massalkhi {et~al.}(2018)Massalkhi, Ag\'{u}ndez, \&
  Cernicharo}]{Massalkhi2018}
Massalkhi, S., Ag\'{u}ndez, M., \& Cernicharo, J. e.~a. 2018, Astron.
  Astrophys., 630, A29

\bibitem[{Matthews {et~al.}(2020)Matthews, Cheng, Harding, Lipparini,
  Stopkowicz, Jagau, Szalay, Gauss, \& Stanton}]{Matthews2020}
Matthews, D.~A., Cheng, L., Harding, M.~E., {et~al.} 2020, J. Chem. Phys., 152,
  214108

\bibitem[{Mauron \& Huggins(2010)}]{Mauron2010}
Mauron, N., \& Huggins, P.~J. 2010, Astron. Astrophys., 513, A31.
\newblock
  \url{https://www.aanda.org/articles/aa/abs/2010/05/aa13970-09/aa13970-09.html}

\bibitem[{Millar(2008)}]{Millar2008}
Millar, T.~J. 2008, Astrophys. Space Sci., 313, 223.
\newblock \url{https://link.springer.com/article/10.1007/s10509-007-9636-z}

\bibitem[{Min {et~al.}(2014)Min, Halfen, \& Ziurys}]{Min2014}
Min, J., Halfen, D.~T., \& Ziurys, L.~M. 2014, Chem. Phys. Lett., 609, 70.
\newblock \url{https://ui.adsabs.harvard.edu/abs/2014CPL...609...70M/abstract}

\bibitem[{Pardo {et~al.}(2021)Pardo, Cabezas, Fonfr\'{i}a, Ag\'{u}ndez,
  Tercero, de~Vicente, Gu\'{e}lin, \& Cernicharo}]{Pardo2021}
Pardo, J.~R., Cabezas, C., Fonfr\'{i}a, J.~P., {et~al.} 2021, Astron.
  Astrophys., 652, L13

\bibitem[{Pardo {et~al.}(2022)Pardo, Cernicharo, Tercero, Cabezas,
  Berm\'{u}dez, Ag\'{u}ndez, Gallego, Tercero, G\'{o}mez-Garrido, de~Vicente,
  \& L\'{o}pez-P\'{e}rez}]{Pardo2022}
Pardo, J.~R., Cernicharo, J., Tercero, B., {et~al.} 2022, Astron. Astrophys.,
  658, A39

\bibitem[{Patel {et~al.}(2011)Patel, Young, Gottlieb, Thaddeus, Wilson, Menten,
  Reid, McCarthy, Cernicharo, He, Br{\"{u}}nken, Trung, \& Keto}]{Patel2011}
Patel, N.~A., Young, K.~H., Gottlieb, C.~A., {et~al.} 2011, Astrophys. J.
  Suppl. Ser., 193, 17.
\newblock \url{https://iopscience.iop.org/article/10.1088/0067-0049/193/1/17
  https://iopscience.iop.org/article/10.1088/0067-0049/193/1/17/meta}

\bibitem[{Petrie(1996)}]{Petrie1996}
Petrie, S. 1996, Mon. Not. R. Astron. Soc., 282, 807.
\newblock \url{https://ui.adsabs.harvard.edu/abs/1996MNRAS.282..807P/abstract}

\bibitem[{Prascher {et~al.}(2011)Prascher, Woon, Peterson, Dunning, \&
  Wilson}]{Prascher2011}
Prascher, B.~P., Woon, D.~E., Peterson, K.~A., Dunning, T.~H., \& Wilson, A.~K.
  2011, Theor. Chem. Acc., 128, 69.
\newblock \url{https://link.springer.com/article/10.1007/s00214-010-0764-0}

\bibitem[{Raghavachari {et~al.}(1989)Raghavachari, Trucks, Pople, \&
  Head-Gordon}]{Raghavachari1989}
Raghavachari, K., Trucks, G.~W., Pople, J.~A., \& Head-Gordon, M. 1989, Chem.
  Phys. Lett., 157, 479.
\newblock
  \url{https://www.sciencedirect.com/science/article/pii/S0009261489873956?via%3Dihub}

\bibitem[{Rehfuss {et~al.}(1988)Rehfuss, Liu, Dinelli, Jagod, Ho, Crofton, \&
  Oka}]{Rehfuss1988}
Rehfuss, B.~D., Liu, D.~J., Dinelli, B.~M., {et~al.} 1988, J. Chem. Phys., 89,
  129

\bibitem[{Sun {et~al.}(2010)Sun, Halfen, Min, Harris, Clouthier, \&
  Ziurys}]{Sun2010}
Sun, M., Halfen, D.~T., Min, J., {et~al.} 2010, J. Chem. Phys., 133, 174301

\bibitem[{Sur {et~al.}(2005)Sur, Sahoo, Chaudhuri, Das, \& Mukherjee}]{Sur2005}
Sur, C., Sahoo, B.~K., Chaudhuri, R.~K., Das, B.~P., \& Mukherjee, D. 2005,
  Eur. Phys. J. D, 32, 25.
\newblock \url{https://link.springer.com/article/10.1140/epjd/e2004-00176-1}

\bibitem[{Thaddeus {et~al.}(1984)Thaddeus, Cummins, \& Linke}]{Thaddeus1984}
Thaddeus, P., Cummins, S.~E., \& Linke, R.~A. 1984, ApJ, 23, L45

\bibitem[{Th{\"{u}}mmel {et~al.}(1989)Th{\"{u}}mmel, Klotz, \&
  Peyerimhoff}]{Thummel1989}
Th{\"{u}}mmel, H., Klotz, R., \& Peyerimhoff, S.~D. 1989, Chem. Phys., 129, 417

\bibitem[{T{\"{o}}rring \& Hoeft(1986)}]{Torring1986}
T{\"{o}}rring, T., \& Hoeft, J. 1986, Chem. Phys. Lett., 126, 477

\bibitem[{Woon(1996)}]{Woon1996}
Woon, D.~E. 1996, Astrophys. J., 456, 602

\bibitem[{Woon \& Dunning(1995)}]{Woon1995}
Woon, D.~E., \& Dunning, T.~H. 1995, J. Chem. Phys, 103, 4572

\bibitem[{Zingsheim {et~al.}(2017)Zingsheim, Martin-Drumel, Thorwirth,
  Schlemmer, Gottlieb, Gauss, \& McCarthy}]{Zingsheim2017}
Zingsheim, O., Martin-Drumel, M.~A., Thorwirth, S., {et~al.} 2017, J. Phys.
  Chem. Lett., 8, 3776

\bibitem[{Ziurys {et~al.}(1995)Ziurys, Apponi, Gu{\'e}lin, \&
  Cernicharo}]{Ziurys1995}
Ziurys, L.~M., Apponi, A.~J., Gu{\'e}lin, M., \& Cernicharo, J. 1995, ApJ, 445,
  L47

\end{thebibliography}
%\bibliographystyle{aasjournal}

\clearpage 
\pagebreak

\begin{deluxetable}{lccC}
    \tablecaption{Laboratory rotational frequencies of MgC$_2$. \label{tab:lab_data}}
\tablewidth{0pt}
\tablehead{
\colhead{Isotopologue}  &  \multicolumn{2}{c}{Transition}  &  \colhead{Frequency} \\
\cline{2-3}
\colhead{}  &  \colhead{$J'_{K_a K_c} - J''_{K_a K_c}$}  &  \colhead{$F'-F''$}  &  \colhead{(MHz)}
}
\startdata
$^{24}$Mg$^{12}$C$_2$ & $1_{01} - 0_{00}$ & $\cdots$ & 20896.090(2) \\
                      & $2_{02} - 1_{01}$ & $\cdots$ &41711.852(4) \\
$^{24}$Mg$^{13}$C$^{12}$C & $1_{01} - 0_{00}$ & $\cdots$ & 20443.702(2) \\
$^{24}$Mg$^{13}$C$_2$ & $1_{01} - 0_{00}$ & $\cdots$ & 20027.207(2) \\
$^{25}$Mg$^{12}$C$_2$ & $1_{01} - 0_{00}$ & $2.5-2.5$ & 20509.780(4)\\
&  &  $3.5-2.5$ & 20512.245(4) \\
&  &  $1.5-2.5$ & 20513.305(4) \\
$^{26}$Mg$^{12}$C$_2$ & $1_{01} - 0_{00}$ & $\cdots$ & 20157.202(2)
\enddata
\tablecomments{Estimated uncertainties are given in parentheses in units of the last digit.}
\end{deluxetable}

%\clearpage
%\pagebreak

\begin{deluxetable*}{l ccCCCcccc}[ht]
\tablecaption{Lines of MgC$_2$ in IRC+10216 \label{tbl-2}}
\tablewidth{0pt}
\tablehead{
\colhead{Isotopologue}  &  \colhead{Transition}   &  \colhead{$\nu_\mathrm{obs}$\tablenotemark{a}}   &  \colhead{$\nu_\mathrm{obs-calc}$}   &  \colhead{$E_{u}/k$}   &  \colhead{$S$}   &  \colhead{$\int{T_\mathrm{A}^{*} dv}$\tablenotemark{b}}   &  \colhead{$\theta_{\mathrm{B}}$}   &  \colhead{$\eta_{\mathrm{B}}$}   &  \colhead{$W$\tablenotemark{c}}   \\
\colhead{}  &  \colhead{$J'_{K_a K_c} - J''_{K_a K_c}$}   &  \colhead{(MHz)}   &  \colhead{(MHz)}   &  \colhead{(K)}   &  \colhead{}   &   \colhead{(mK~\kms)}   &  \colhead{$(\arcsec)$}   &  \colhead{}   &  \colhead{(mK~\kms)}       
}
\startdata
$^{24}$MgC$_2$   &   $2_{02}-1_{01}$   &   $41711.82\pm0.02$  &   -0.034   &   3.00   &   2.00                      &     $221\pm10$   &   42.5 &   0.53   &   $1254\pm138$ \\
                 &   $4_{04}-3_{03}$ & $82787.94\pm0.02$  &   0.015   &   9.97   &   3.99       &     $362\pm10$   &   29.7 &   0.853   &   $841\pm87$ \\
                 &   $4_{23}-3_{22}$ & $83511.00\pm0.15$  &   -0.096   &  17.98   &   3.00       &     $108\pm11$   &   29.5 &   0.853   &   $249\pm36$ \\
                 &   $4_{22}-3_{21}$ & $84304.13\pm0.04$  &   0.105   &  18.04   &   3.00       &     $149\pm9$   &   29.2 &   0.853   &   $340\pm40$ \\
                 &   $5_{05}-4_{04}$ & $102907.74\pm0.10$  &   0.137   &  14.91   &   4.98      &     $248\pm8$   &   23.9 &   0.853   &   $476\pm50$ \\
                 &   $5_{42}-4_{41}$\tablenotemark{d} &  $104610.76\pm0.30$  &   -0.463   &  46.88   &   1.80     &     $35\pm8$   &   23.5 &   0.853   &   $65\pm16$ \\
                 &   $5_{41}-4_{40}$\tablenotemark{d} & $104610.76\pm0.30$  &   -0.762   &  46.88   &   1.80      &     $35\pm8$   &   23.5 &   0.853   &   $65\pm16$ \\
                 &   $5_{23}-4_{22}$ & $105852.05\pm0.15$  &   -0.338   &  23.12   &   4.20      &     $164\pm9$   &   23.2 &   0.853   &   $308\pm35$ \\
                 &   $6_{06}-5_{05}$ & $122684.00\pm0.50$  &   -0.084   &  20.80   &   5.97      &     $251\pm30$   &   20.1 &   0.785   &   $463\pm89$ \\
                 &   $7_{07}-6_{06}$ & $142108.88\pm0.10$  &   -0.077   &  27.62   &   6.96      &     $83\pm6$   &   17.3 &   0.785   &   $141\pm23$ \\
                 &   $7_{25}-6_{24}$ & $149761.18\pm0.20$  &   -0.189   &  36.43   &   6.43      &     $38\pm7$   &   16.4 &   0.785   &   $63\pm15$ \\
$^{25}$MgC$_2$   &   $2_{02}-1_{01}$   &   $40949.33\pm0.20$  &   -0.267   &   3.00   &   2.00      &     $34\pm5$   &   42.5 &   0.53   &   $193\pm34$ \\
$^{26}$MgC$_2$   &   $2_{02}-1_{01}$   &   $40245.20\pm0.20$  &   -0.085   &   2.90   &   2.00   &     $46\pm8$   &   42.5 &   0.53   &   $261\pm52$ \\
                 &   $4_{04}-3_{03}$ & $79942.90\pm0.30$  &   0.171   &   9.62   &   3.99       &     $41\pm13$   &   29.7 &   0.853   &   $99\pm33$ \\
                 &   $5_{05}-4_{04}$ & $99428.70\pm0.40$  &   -0.138   &  14.40   &   4.98       &     $45\pm10$   &   29.7 &   0.853   &   $89\pm22$ 
\enddata
\tablecomments{Unless otherwise noted, estimated uncertainties are $1\sigma$. Frequencies subtracted from those observed are calculated from the constants in Table~\ref{tbl-3}. $E_{u}$ is the energy of the upper state of the transition, and $S$ is the asymmetric rotor line strength.  $\mathrm{\alpha_{J2000}=09^{h}47^{m}57^{s}{.}36,\delta_{J2000}=+13^{\circ}16\arcmin44{.}{\arcsec}4}$.}
\tablenotetext{a}{On the assumption of $v_\mathrm{LSR}=-26.5$~\kms{}.}
\tablenotetext{b}{Derived from least-squares fits of shell profiles to the spectra shown in Figs.~\ref{fig:astro_main} and \ref{fig:astro_isotope}; in some cases the expansion velocity was fixed to $14.5$~\kms.}
\tablenotetext{c}{$W=\int{T_\mathrm{A}^{*} dv}/\eta_\mathrm{B}f_\mathrm{D}$, where $f_\mathrm{D} = \theta_\mathrm{S}^2/(\theta_\mathrm{S}^2 + \theta_\mathrm{B}^2)$ and the source diameter $\theta_\mathrm{S} = 30\arcsec$. Errors include calibration uncertainties of $10\%$ at 7~mm, $10\%$ at 3~mm, and $15\%$ at 2~mm, as well as an assumed error of 1~D on the dipole moment, added in quadrature.}
\tablenotetext{d}{Unresolved doublet with each component assumed to lie at the same frequency and possess half the intensity.}
%\tablerefs{(1) This work; (2) TK.}
\end{deluxetable*}

%\clearpage
%\pagebreak

\begin{deluxetable}{l C C C}
\tablewidth{0pt}
\tablecaption{Spectroscopic Constants of MgC$_2$ \label{tbl-3}}
\tablehead{
\colhead{Constant}  &  \colhead{$^{24}$MgC$_2$}   &  \colhead{$^{25}$MgC$_2$}  &  \colhead{$^{26}$MgC$_2$}      
}
%\tablehead{
%\colhead{Constant}  &  \multicolumn2c{$^{24}$MgC$_2$}   &  \multicolumn2c{$^{25}$MgC$_2$}  &  %\multicolumn2c{$^{26}$MgC$_2$}            
%}
%\decimals
\startdata
\multicolumn{4}{l}{Rotational and centrifugal distortion constants (MHz):} \\
$A$\dotfill  &  51900(1)  &  51899.8\tablenotemark{b}  &  51897.6\tablenotemark{b} \\
$(B+C)/2$\dotfill    &  10448.0740(9)  & 10255.8294(12)  &  10078.601(1)  \\
$(B-C)/4$\dotfill    &    526.033(14)  &  507.046\tablenotemark{b}  &  490.890(61) \\
$10^{3}\Delta_{J}$\dotfill  &  14.454(101)   &  \cdots  &  \cdots \\
$10^{3}\Delta_{JK}$\dotfill  & 245.12(169)   &  \cdots  &  \cdots \\
$10^{3}\Delta_{K}$\dotfill  &   -37.84\tablenotemark{a}  &  \cdots   & \cdots \\
$10^{3}\delta_{J}$\dotfill  &  0.369(168)   &  \cdots  & \cdots \\
$10^{3}\delta_{K}$\dotfill  & 148.03\tablenotemark{a}  &  \cdots   & \cdots \\
\multicolumn{4}{l}{Electric quadrupole coupling constants (MHz):}  \\
$\chi_{aa}$\dotfill   &   \cdots  & -11.75(2)  &  \cdots \\
$\chi_{bb}$\dotfill   &   \cdots  &   4.8(10)  &  \cdots \\
\hline
$\sigma_\mathrm{nrms}$\dotfill  & 1.37  &  0.29   & 0.40 \\
\hline
\multicolumn{4}{l}{Inertial defect (amu-\AA$^{2}$):}  \\
$\Delta$\dotfill  &  0.1035(3)  &   0.1037   &  0.125(1)
\enddata
\tablecomments{Constants are derived from a least-squares fit to the rotational transitions in Tables~\ref{tab:lab_data} and \ref{tbl-2}.  The $1\sigma$ uncertainties are indicated in  parentheses, with the least significant digit corresponding to the least significant digit of the value.  Parameters for which no errors are indicated have been fixed to the theoretical values.}
\tablenotetext{a}{Second-order vibrational perturbation theory (VPT2) value at the CCSD(T)/cc-pVTZ level of theory.}
\tablenotetext{b}{Fixed to the experimental values derived from $^{24}$MgC$_2$: the equilibrium isotope shift was taken from the semi-experimental equilibrium structure, and the zero-point correction from the variational calculations (Appendix~\ref{sec:calc}).}
\end{deluxetable}

\pagebreak

\begin{deluxetable}{cLL}
    \tablecaption{The equilibrium geometry of MgC$_2$. \label{tab:geometry}}
%\tablewidth{0pt}
\tablehead{
\colhead{Basis set}  &  \colhead{$r_\mathrm{CC}$}  &  \colhead{$r_\mathrm{Mg-C_2}$}
}
\startdata
D & 1.29422 & 1.96504 \\
T & 1.27545 & 1.92117 \\
Q & 1.27175 & 1.91073 \\
5 & 1.27082 & 1.90741 \\
CBS & 1.27068 & 1.90673 \\ 
\hline
$r_\mathrm{se}$ & 1.2706(7) & 1.9066(1)
\enddata
\tablecomments{The CCSD(T)/cc-pCV$X$Z bond lengths (in $\mathrm{\AA}$) are shown for $X$ = D--5, followed by the complete basis set (CBS) extrapolation. The semi-experimental equilibrium geometry ($r_{\mathrm{se}}$) was determined as discussed in the text. $r_{\mathrm{CC}}$ is the C-C distance and $r_{\mathrm{Mg-C_2}}$ is the distance between Mg and the CC center-of-mass.}
\end{deluxetable}

\pagebreak

\begin{figure}[h]
\centering 
\includegraphics[width=3.4in]{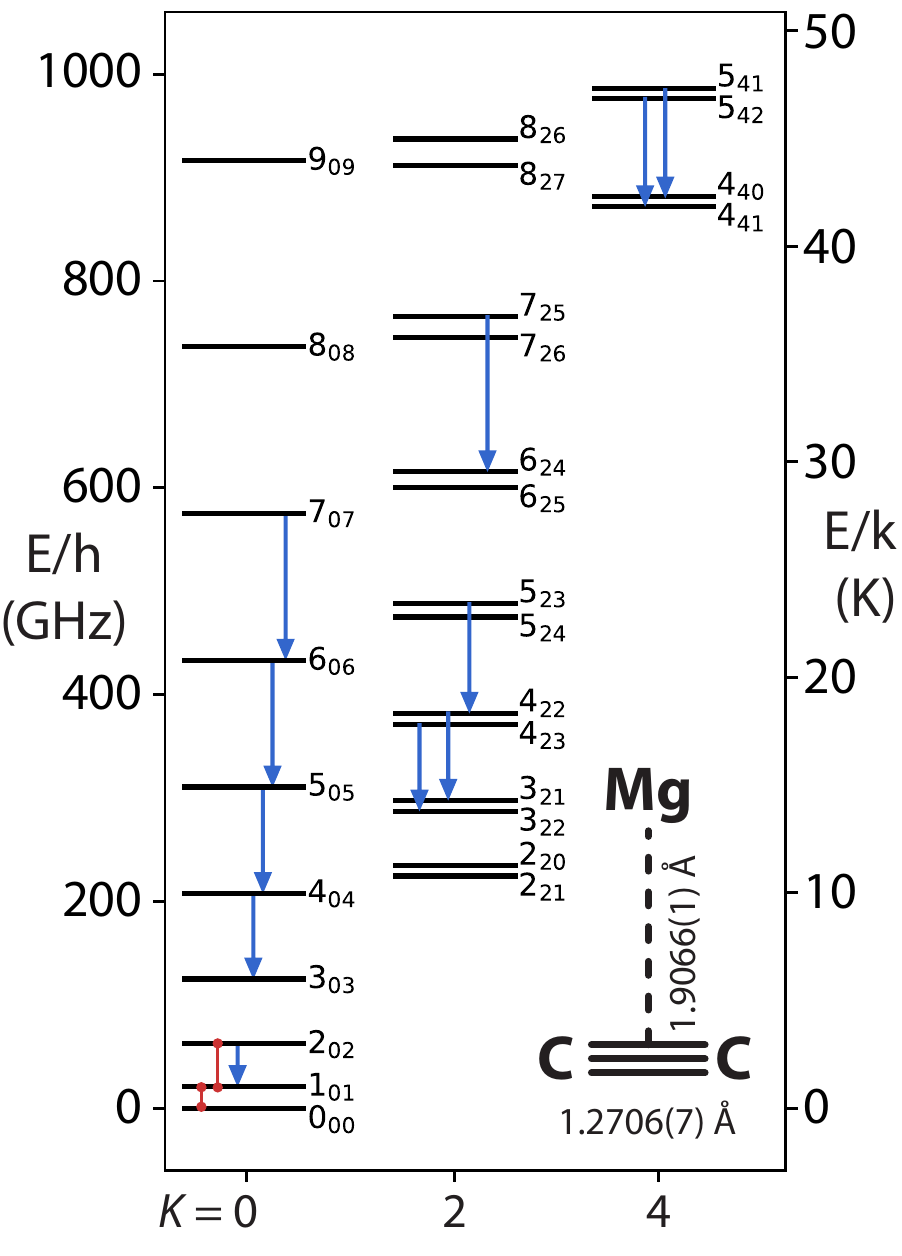}
\caption{\label{fig:rotation_levels} Lower rotational energy levels of MgC$_2$. Red lines indicate the observed laboratory transitions and blue arrows the astronomical transitions. The inset shows the semi-experimental equilibrium structure.}
\end{figure}

\pagebreak

\pagebreak

\pagebreak

\pagebreak

\end{document}